\documentclass[twocolumn]{aastex631}

\begin{document}

\title{Tracing the Winds: A Uniform Interpretation of Helium Escape in Exoplanets from Archival Spectroscopic Observations}

\author[0000-0003-0473-6931]{Patrick McCreery}
\affiliation{Department of Physics and Astronomy, Johns Hopkins University, Baltimore, MD 21218, USA}
\email{pmccree2@jhu.edu}

\author[0000-0002-2248-3838]{Leonardo A. Dos Santos}
\affiliation{Space Telescope Science Institute, 3700 San Martin Drive, Baltimore, MD 21218, USA}
\affiliation{Department of Physics and Astronomy, Johns Hopkins University, Baltimore, MD 21218, USA}

\author[0000-0001-9513-1449]{N\'estor Espinoza}
\affiliation{Space Telescope Science Institute, 3700 San Martin Drive, Baltimore, MD 21218, USA}
\affiliation{Department of Physics and Astronomy, Johns Hopkins University, Baltimore, MD 21218, USA}

\author[0000-0002-1199-9759]{Romain Allart}
\altaffiliation{Trottier Postdoctoral Fellow}
\affiliation{D\'epartement de Physique, Institut Trottier de Recherche sur les Exoplan\`etes, Universit\'e de Montr\'eal, Montr\'eal, Qu\'ebec, H3T 1J4, Canada}

\author[0000-0002-4207-6615]{James Kirk}
\affiliation{Department of Physics, Imperial College London, Prince Consort Road, London, SW7 2AZ, UK}

\begin{abstract}

Over the past decade, observations of evaporating exoplanets have become increasingly common, driven by the discovery of the near-infrared helium-triplet line as a powerful probe of atmospheric escape. This process significantly influences the evolution of exoplanets, particularly those smaller than Jupiter. Both theoretical and observational studies have aimed to determine how efficiently exoplanets convert their host star’s X-ray and ultraviolet (XUV) radiation into atmospheric mass loss. In this study, we employ the open-source atmospheric escape model \textsf{p-winds} to systematically analyze all publicly available helium triplet spectroscopic detections related to exoplanetary atmospheric escape. Our findings indicate that the retrieved outflows strongly depend on the ratio of XUV flux to planetary density ($F_{\text{XUV}}/\rho_p$), supporting the theoretical framework of energy-limited mass loss. We constrain population-level photoevaporative efficiencies to $0.34 \pm 0.13$ and $0.75 \pm 0.21$ for hydrogen-helium fractions of $0.90$ and $0.99$, respectively. These results offer new insights into exoplanetary atmospheric evolution and will aid future studies on exoplanet population demographics.

\end{abstract}


\section{Introduction} \label{sec:intro}

Exoplanet surveys have revealed that a large fraction of exoplanets orbit at short periods, ranging from hours to a few days \citep[e.g.,][]{batalha2013, morton2016, Zhu2021}. One unresolved question posed during the first years of discoveries was whether hot gas giants were able to retain their atmospheres given their proximity to their host stars \citep[][]{Guillot1996}. The general consensus is that photoevaporation driven by extreme ultraviolet (XUV) irradiation \citep[e.g.,][]{Lammer2003, Lecavelier2007, Davis2009, Salz2016b} in combination with dynamical effects \citep[][]{Owen2018, Vissapragada2022b} are the main engines for the onset of atmospheric escape in hot gas giants. It is now understood that hot Jovian planets are resilient against evaporation due to their higher gravity \citep[e.g.,][]{Lazovik2023, Owen2018}, but the same cannot be said about smaller exoplanets \citep[e.g.,][]{Fulton2017}. 

Atmospheric escape has an important role in changing the radii of sub-Jovian planets \citep[e.g.,][]{Owen2013, Kurosaki2014, Lundkvist2016}, but we lack observational constraints on how efficiently XUV irradiation is converted into an outflow \citep[e.g.,][]{Shematovich2014, Krenn2021, Caldiroli2022}. Namely, the most important astrophysical parameter that sets the evolution course of an exoplanet is its mass-loss rate. With the discovery of the metastable helium (He) triplet at 1.083~$\mu$m as a tracer for atmospheric escape \citep{Seager2000, Oklopcic2018, Allart2018, Mansfield2018, Nortmann2018, spake2018}, we have begun an era where observational surveys for evaporating planets are feasible \citep[e.g.,][]{Lampon2021, Vissapragada2022b, MZhang2023, Lampon2023, Allart2023, Masson24, Orell24}. These surveys provide us with an opportunity to measure mass-loss rates that will inform detailed models of exoplanetary evolution, which in turn will be crucial for the search of atmospheres of rocky, potentially habitable, exoplanets \citep[e.g.,][]{Zahnle2017, Zieba2023, KTotton2023, Teixeira2024}. 

According to \citet{DSantos2023a}, at least 14 exoplanets have been shown to have metastable helium outflows detected with high spectral resolution or narrow-band photometry, as well as five with tentative detections. Since then, new detections have been reported, such as those contained in \citet{Zhang2023}, \citet{PGonzalez2023} and \citet{BArufe2023} for HAT-P-32~b, TOI-1268~b and HAT-P-67~b, respectively.

Early efforts in searching for demographic trends in evaporating exoplanets suggested that the in-transit helium absorption level, measured in units of scale height, is roughly proportional to the XUV irradiation that the planet receives \citep[e.g.,][]{Nortmann2018, DSantos2020a}. However, more recent results have shown that this trend is less clear with larger samples \citep[e.g.,][]{Kirk2022, OMiquel2022, Tyler2023}. In theory, a clearer trend should arise in the measured mass-loss rates in function of XUV irradiation divided by the planetary density, which is predicted by the energy-limited formulation \citep{Watson1981, Lecavelier2007, Salz2016a}. \citet{Vissapragada2022b} studied a sample of seven evaporating gas giants observed in narrow-band photometry, and found that five of them follow this trend. A study of spectroscopic detections and non-detections using the SPIRou instrument found no significant trend between the XUV irradiation and inferred mass-loss rates \citep{Allart2023}. Finally, a study by \citet{MZhang2023} reported on a statistically significant trend, although it relies on equivalent width measurements instead of mass loss estimates. 

We present here a uniform analysis of all public metastable helium observations with the goal of investigating the existence of the theoretical relationship between mass-loss rate and XUV irradiation divided by the planetary density. Performing a uniform analysis will remove any discrepancies due to the variety of modeling frameworks used in previous studies, allowing us to investigate true population trends.

This manuscript has the following structure: in Section \ref{sec:methods}, we describe how we interpreted the transmission spectra with the {\sf p-winds} code and Gaussian processes. Section \ref{sec:archival-data} briefly describes the archival observations. In Section \ref{sec:results}, we discuss the main results of our investigation for each exoplanet in the sample, and Section \ref{sec:interpretation} contains the interpretation of these results. We close this manuscript with the conclusions in Section \ref{sec:conclusions}.

\section{Methods}\label{sec:methods}

The purpose of this study is to measure the mass-loss rates of all exoplanets with detections of escaping helium (at the time of the analysis) in high-resolution transmission spectroscopy using a uniform modeling framework across all observations. However, given that the observations are sourced from different observatories, we must address data systematics and the modeling of mass-loss rates to ensure a uniform analysis. 

\subsection{Data Sample and Analysis}\label{sec:data}

We retrieved archival exoplanet transmission spectra available in online databases or by private communication, outlined in Table \ref{tab:sample}. We note that the transmission spectra may be reduced in different ways by the different teams associated with each observation. While reducing the data under a uniform framework would be the most ideal way to approach this analysis, we do not expect this to have a large impact on our analysis and the work to re-reduce data is outside of this project's scope. In situations where there is more than one observation of a single exoplanet in helium with either Keck or CAO, we use the dataset more readily available. However, \citet{Kirk2020} verified that for WASP-107b, the interpretation of the helium-triplet signature was not significantly different between the two observatories.

For 8 of the 12 spectra, we had access to the one-dimensional transmission spectrum, which required no further reduction. However, for the planets TOI-560b, TOI-1430b, TOI-1683b, and TOI-2076b, we reduced the two-dimensional transmission spectra in the stellar rest frame to obtain the one-dimensional spectra in the planetary rest frame. \cite{Zhang_2023} discusses the full extent of reducing the spectra and is the process we follow here. For all of the spectra, we ensured that the wavelengths were all measured in-air, via a correction if necessary.

The spectral energy distributions (SEDs) used are shown in Table \ref{tab:sample}. HD 189733, HD 209458, TOI-1430, and TOI-1683 had published SEDs that were used in lieu of proxies. For the stars without SED observations, we identified best-fit proxies to use in the modeling process. The proxy identification process consists of matching the properties of the exoplanet host star to those of a star that have an EUV SED from the MUSCLES database (\citet{musclesI}, \citet{musclesII}, \citet{musclesIII}, \citet{musclesIV}, \citet{musclesV}, \citet{France2016}). We utilize the spectral type and activity index or age to match stellar hosts with SED proxies in the database, then scale the archival SED to the radius of the stellar host and the semi-major axis of the planet to yield a best estimate of the SED incident on the planet.

\begin{deluxetable*}{lllll}[ht!]
\centering
\tablecaption{Sample of exoplanets with archival observations of helium escape.\label{tab:sample}}
\tablecomments{SED proxies from the MUSCLES survey \citep{France2016} combine X-rays and UV spectra with Differential Emission Measure to determine the unobservable extreme-UV (EUV) spectrum. The SEDs from the X-exoplanets database \citep{SForcada2022} also use X-rays and UV observations, but the EUV spectrum is calculated using a coronal model.}
\tablewidth{0pt}
\tablehead{
\colhead{Object} & \colhead{Observatory/Instrument} & \colhead{Data Reference} & \colhead{SED proxy} & \colhead{SED reference} \\
}
\startdata
GJ-3470b  & CAO/CARMENES & \cite{Lampon2021}    & GJ 176 & \citet{France2016} \\
HAT-P-11b & CAO/CARMENES & \cite{Allart2018}    & HD 40307 & \citet{France2016} \\
HD 189733b & CAO/CARMENES & \cite{Salz2018}    & HD 189733 & \citet{SForcada2022} \\
HD 209458b & CAO/CARMENES & \cite{Lampon2020}    & HD 209458 & \citet{SForcada2022} \\
WASP-52b  & Keck II/NIRSPEC & \cite{Kirk2022}   & eps Eri & \citet{France2016} \\
WASP-69b  & CAO/CARMENES & \cite{Nortmann2018}  & HD 40307 & \citet{France2016}  \\
WASP-107b & Keck II/NIRSPEC & \cite{Kirk2020}   & HD 85512 & \citet{France2016}  \\
WASP-177b & Keck II/NIRSPEC & \cite{Kirk2022}   & HD 40307 & \citet{France2016}  \\
TOI-560b  & Keck II/NIRSPEC & \cite{Zhang_2023} & eps Eri & \citet{France2016}  \\
TOI-1430b & Keck II/NIRSPEC & \cite{Zhang_2023} & TOI-1430 & \citet{Zhang_2023} \\
TOI-1683b & Keck II/NIRSPEC & \cite{Zhang_2023} & TOI-1683 & \citet{Zhang_2023} \\
TOI-2076b & Keck II/NIRSPEC & \cite{Zhang_2023} & eps Eri & \citet{France2016}  \\ 
\enddata
\end{deluxetable*}

We note that using proxies in this analysis introduces uncertainty in the true incident irradiation on the orbiting planets; propagating this error is difficult to do and does not significantly contribute to systematic errors \citep{Zhang24}. However, further study of exoplanetary host stars is critical to our understanding of mass loss, and thus evolution of extrasolar systems. 

Detections of the helium-triplet feature are more information-rich than non-detections, as we can place constraints on the mass-loss parameters. However, non-detections provide an upper-bound on the mass-loss rate, which are especially useful to constrain the variability of the helium-triplet line and to add constraints to population analyses. The difficulty of bounding the mass-loss rates, recovering useful uncertainties, and interpreting the results led us to remove them from this analysis. \citet{DosSantos_2021} contains references to all non-detections reported before that publication, and other non-detections have since been published \citep[e.g.,][]{Bennett_2023, Vissapragada_2024, Alam_2024}. 

\subsection{Instrumental Systematics Model}\label{sec: GP}

\begin{figure}[ht!]
    \centering
    \includegraphics[width=.45\textwidth]{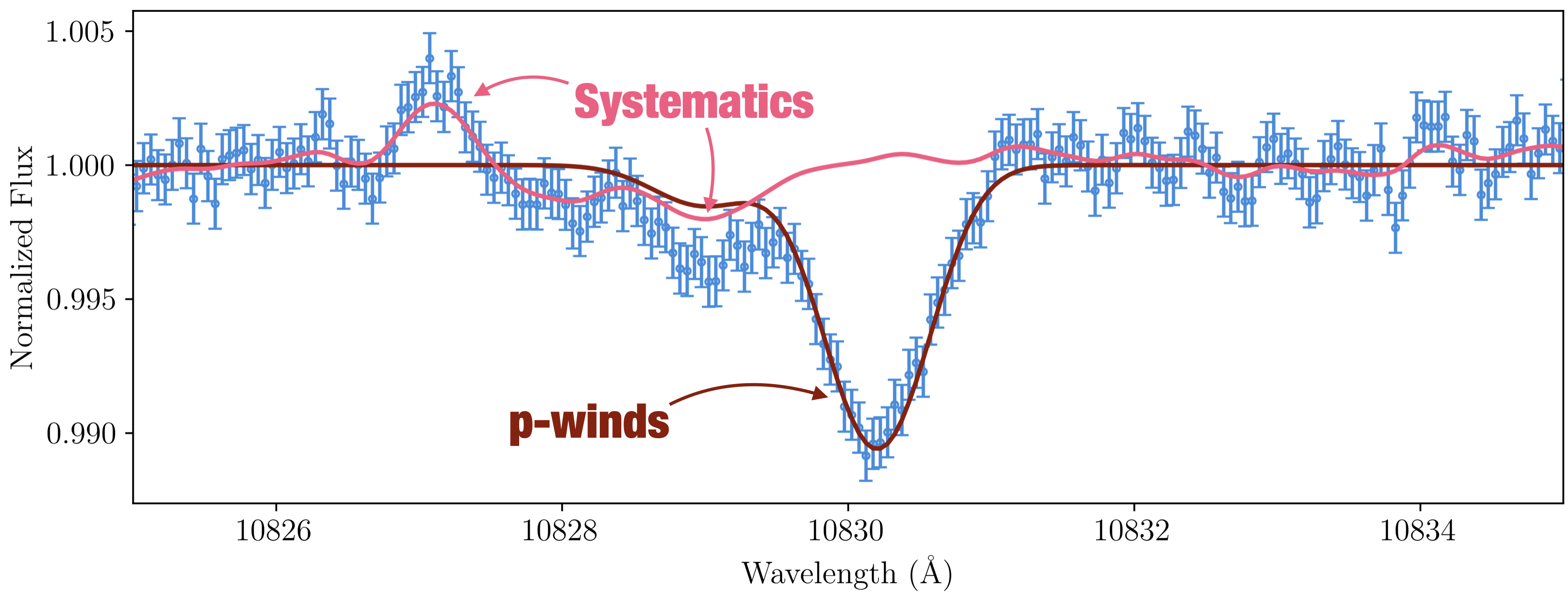}
    \caption{Overview of the systematics present in an observed transmission spectrum. In pink is a best fit GP to the systematics in the sample, and dark red is a best fit of the helium-triplet using \textsf{p-winds}.}
    \label{fig:hd189systematics}
\end{figure}

Although the one-dimensional transmission spectra are reduced, there are obvious sources of correlated noise in the data that must be accounted for in the modeling framework (see Figure \ref{fig:hd189systematics}). Unaddressed correlated noise in transmission spectra has been shown to be a potential source of bias, causing inferences to extract features in spectra that are not present \citep{ih:2021}. Furthermore, removing correlated noise aids in creating a uniform sample of helium triplet observations that are taken from distinct observatories under distinct conditions. We perform a simultaneous fit of the mass-loss and systematics models, which will be discussed in Section \ref{sec:pwinds}.

We model correlated noise in our spectra via Gaussian Processes (GPs) -- a non-parametric approach that allows us to capture a wide range of different manifestations of correlated noise and underlying patterns in data while accounting for the uncertainty in these fits. Figure \ref{fig:hd189systematics} displays an example of the need to model correlated noise, as well as the impacts correlated noise could have on retrievals. Without an understanding of the correlated noise, a mass-loss model is subject to fitting noise, not the true signal. The use of Gaussian Processes in the context of exoplanet transmission spectrum analyses are becoming more commonly used, including the publication of \citet{guilluy24} during this work's analysis.

GPs define a distribution over functions, where any finite set of function values follows a multivariate Gaussian distribution. This means that GPs provide a way to estimate the possible functions that could generate a given dataset, while considering different levels of smoothness or uncertainty. We use the package \textsf{celerite} \citep{celerite} to implement GPs into our model.

The heart of GPs lie in their kernel function, or covariance function, which encodes assumptions about the relationships between the data points. By adjusting kernel parameters, you can control the trade-off between fitting the observed data and allowing for flexibility in the function's behavior. For this study, we use the Matérn-3/2 kernel which emphasizes smoothness in the functions drawn while accounting for sharp changes between data points that we observe in transmission spectra. The GPs contributes 2 tunable parameters when fitting the GP models to correlated noise.

In an effort to minimize any over-fitting of the Gaussian process, we perform a GP fit outside of the helium-triplet window and use this initial fit to place constraints on the GP during the simultaneous instrumental and mass-loss model fit. This means that when using the Gaussian process inside of the helium-triplet window, we are confident that we are minimizing the component of the GP that is fitting the helium-triplet feature instead of systematics. Provided the biases that come along with fitting the helium feature by a mass-loss model alone, we propose this to be a more robust method of retrieving mass-loss parameters.

\subsection{Atmospheric Escape Model}\label{sec:pwinds}

We utilize the \textsf{p-winds} code version $1.4.6$ \citep{10.5281/zenodo.10419485,pwinds} to produce atmospheric escape models. \textsf{p-winds} is based on the Parker-wind approximation \citep{Parker58} applied for exoplanet outflows from the upper atmosphere \citep{Oklopcic2018}. This code has been used extensively in the literature \citep[e.g.,][]{Vissapragada2022b, Kirk2022, Guilluy2023, Allart2023, BArufe2023, Tyler2024}, so we describe the framework only briefly here. As fixed input, \textsf{p-winds} takes the stellar and planetary parameters, as well as the stellar host spectral energy distribution. The four free parameters of this framework are: the planetary mass-loss rate $\dot{m}$, the outflow temperature $T$, bulk line-of-sight velocity of the outflow $v$, and the helium abundance. For the latter, we chose to either assume that the planets have solar helium fraction (10\%), or that their helium fraction is 1\%. The reason for the second choice is that recent studies using self-consistent hydrodynamic models to fit metastable helium observations suggest that the outflows of evaporating exoplanets have low helium abundance, with a He/H number fraction close to 1/99 \citep{Salz2016b, Rumenskikh2022, Yan2024}.

The code assumes that the sound speed of the outflow is constant with distance from the planet, which allows for a significant simplification of the Euler equations that describe a hydrodynamic outflow. First, {\sf p-winds} calculates the hydrogen ionization fraction as a function of radius, followed by the density and velocity profiles of the outflow. Secondly, it calculates the distribution of metastable helium, and thirdly it uses ray-tracing and radiative transfer algorithms to calculate the average in-transit absorption spectrum. We refer the reader to \citet{Oklopcic2018}, \citet{Lampon2020} and \citet{pwinds} for more details on the modeling framework.

We take the stellar and planetary parameters from the NASA Planetary Systems Composite Table \citep{10.26133/NEA13}. The stellar high-energy SEDs are unobserved for most of the stars in the sample, so for those we adopt proxy SEDs or use their SEDs inferred from a combination of X-rays + UV observations and modeling (see Section \ref{sec:data} and Table \ref{tab:sample}).

Simultaneously fitting the helium triplet feature and correlated noise, we are able account for the dominant features in the population's transmission spectra, ensuring quality fits regardless of instrument or object.

\subsection{Parameter Fitting -- Nested Sampling}

With an individual model discussed, we turn to a discussion of parameter space searches and methods for best fitting the \textsf{p-winds} + GP model to observational data.

We utilize nested sampling to constrain the five parameters’ posterior distributions (2 and 3 contributed by the GP and \textsf{p-winds}, respectively). Nested sampling is a robust method of generating posterior samples in complex parameter spaces and provides accurate estimations of the marginal likelihood (Bayesian evidence). Nested samplers use live points from a previously defined prior distribution and replace the lowest likelihood live point with a new live point that has a larger likelihood. Convergence is stopped when the change in the Bayesian inference between iterations crosses a user-defined tolerance, which we set as 0.010 for this work. Dynamic nested sampling is further employed, which dynamically changes the number of live points, allowing the sampler to improve on the distribution of samples in the prior space. We deploy the package \textsf{dynesty} to implement nested sampling into our analysis.

Nested samplers thoroughly search the prior space to estimate evidences, meaning the sampling routine can find true solutions to complex, potentially multi-modal posterior distributions. For this specific application, we use nested sampling to find a best fit and define confidence intervals for our model given the transmission spectra in our population sample. Finally, to retrieve posterior distributions for the model parameters, we resample the set of points according to the posterior weights provided by \textsf{dynesty}.

Bringing these three utilities together, we use \textsf{p-winds} to fit the helium triplet observation, Gaussian processes to handle systematics, and nested sampling to adjust these model parameters and investigate their posterior distributions. The nested sampling routine provides log-likelihood values that we can use to provide a best-fit to the transmission spectrum (quantifying the likelihood of obtaining the observed data given the model parameters).

Across a sample of exoplanets, this model becomes computationally expensive. We utilize \textsf{ray} to parallelize the model fits to save computational time. 

\section{Archival Observations}\label{sec:archival-data}

As the number of metastable helium observations continues to grow, the characteristics of exoplanets studied becomes more varied and requires attention. This study utilizes observational data of hot Jupiters, warm Neptunes, and mini Neptunes. Here we describe the population and previous studies of each object.

\subsection{Jupiters}

At the start of this analysis, our population size with publicly available helium-triplet observations is twelve from CAO/CARMENES and Keck II/NIRSPEC. This population includes a diverse range of planet types and atmospheric escape intensities, making it well-suited to provide insights into energy-limited mass loss.  We utilize the archival observations that are most readily available and easily obtained.

\begin{itemize}
\item HD 189733b is a hot Jupiter orbiting the highly active K dwarf HD 189733 that highly irradiates the planet, which should cause substantial atmospheric mass loss. \citet{Salz2018} observed this exoplanet with the CARMENES spectrograph, finding that the prominent layers where the planet is losing its atmosphere is not traced by the helium triplet. \citet{Zhang2022} observed this exoplanet with Keck II/NIRSPEC, identifying significant stellar XUV variability that results in variable helium absorption. Other observations include those from \citet{Allart2023} and \citet{masson2024} with SPIRou and \citet{guilluy2020} using GIARPS.

\item HD 209458b is one of the most studied hot Jupiter exoplanets due to the system being bright and nearby. HD 209458 is a G star that is relatively inactive and only moderately irradiates HD 209458b with XUV flux. \citet{floriano2019} and \citet{Lampon2020} utilized CARMENES to observe the helium-triplet feature, where the latter study constrained the thermospheric structure and mass-loss rate of the planet. Observations of Lyman-$\alpha$ have also proven important for investigating the evolution of the planet. \citet{Lampon2020} and references therein outline the extensive studies of HD 209458b and its high-efficiency escape. Other observations include those from \citet{Masson24} with SPIRou.

\item WASP-52b is an inflated hot Jupiter orbiting a K dwarf that is both young and active. \citet{Kirk2022} observed this exoplanet with NIRSPEC, noting the low surface gravity of the planet as being a motivating factor for atmospheric escape study. In conjunction with WASP-69b, \citet{Vissapragada2020} utilized the Wide-field Infrared Camera on the 200-inch Hale Telescope at Palomar Observatory to prove that ultra-narrowband photometry is a viable observing technique to quantify helium triplet absorption. Other observations include those from \citet{Allart2023} with SPIRou.

\item WASP-69b is an extended hot Jupiter orbiting a K type star that highly irradiates WASP-69b. \citet{Nortmann2018} observed this exoplanet with the CARMENES spectrograph, finding post-transit absorption pointing to the existence of an extended evaporating tail. Other observations include those from \citet{Allart2023} and \citet{masson2024} with SPIRou.

\item WASP-107b is an extremely low density warm Jupiter orbiting a K star and is the host to the first detection of metastable helium in an exoplanet atmosphere \citep{spake2018} by utilizing \textit{HST}. Previous observations include that with \textit{HST} and the Wide-Field Camera 3 (WFC3) instrument, the CARMENES spectrograph on the 3.5 m telescope at the Calar Alto Observatory, and the Keck II/NIRSPEC spectrograph \citep{Allart2019, Kirk2020}. \citet{Allart2019} and \citet{Spake_2021} discovered and characterized a helium tail. 

\item WASP-177b is a hot Jupiter orbiting an old K dwarf. \citet{Kirk2022} observed this exoplanet with NIRSPEC and found a tentative detection of metastable helium, with similar motivations of a low surface gravity. WASP-177b is of specific interest due to it being on the upper edge of the mass-radius distribution \citep{turner2019}.

\end{itemize}

\subsection{Neptunes}

\begin{itemize}
\item GJ-3470b is a warm Neptune orbiting an M type star. Located close to the sub-Neptune desert, this object has been the subject of many studies indicating the presence of a large hydrogen exosphere with a high mass-loss rate \citep{bourrier2018}. \citet{Ninan_2020} detected helium using the Habitable Zone Planet Finder spectrograph on the Hobby-Eberly Telescope. \citet{Palle2020} also observed GJ-3470b with CARMENES, constraining the mass-loss rate, outflow temperature, and the stellar and planetary parameters. Other observations include those from \citet{Allart2023} and \citet{masson2024} with SPIRou.

\item HAT-P-11b is a warm Neptune orbiting a K type star with a predicted low enough mass-loss rate such that the planet would have only lost a small portion of its mass over its history. \citet{Allart2018} and \citet{Mansfield2018} both detected a helium signature using HST's WFC3 and CARMENES, respectively. This would be the first instance in which independent detections of the helium triplet were made from both ground and space. Other observations include those from \citet{Allart2023} and \citet{masson2024} with SPIRou. 

\end{itemize}

\subsection{mini-Neptunes}

TOI-560b, TOI-1430b, TOI-1683b, and TOI-2076b are all young mini-Neptunes in systems with K type host stars that are younger than $\sim$ 600 Myr with varying incident XUV fluxes. The mass estimates of each of these objects are not well constrained, leaving their planetary densities uncertain. As outlined in \citet{Zhang_2023}, the masses of the exoplanets, besides TOI-560b, are derived from a mass-radius relation \citep{Wolfgang_2016}. The ages of the systems were obtained using the host star rotation period. \citet{Zhang_2022} and \citet{Zhang_2023} observed the young planets with NIRSPEC.

\citet{Zhang_2023} outlines that each of these four mini-Neptunes have primordial atmospheres that are evaporating and will plausibly become super-Earths. In our sample and given the proxies used, TOI-1683b is the most irradiated exoplanet in the XUV by its host star, providing an interesting comparison of mass-loss estimates.

\section{Results and discussion}\label{sec:results}

In this Section, we lay out the results of our mass-loss investigations and discuss these results in the context of previous interpretations of the same or similar data. In the context of comparing atmospheric escape retrievals, \citet{Lampon2020} developed a one-dimensional hydrodynamic model paired with a non-LTE model, and will henceforth be referred to as the \citet{Lampon2020} model. We also compare our results with those obtained with the EVE code \citep{bourrier15, Allart2018}, which is a kinematic model with charge exchange. The paper outlining the Parker wind model used in this analysis will be referred to as the \citet{pwinds} model. 

In Table \ref{tab:ObsResults}, we report the bounds on the peak excess helium-triplet absorption. We select the 95\% and 99.7\% upper-bound on peak absorption to demonstrate the maximum expected depth of the observed helium-triplet signature for each exoplanet. Table \ref{tab:Results9099} outlines the results of our modeling efforts including the mass-loss rate, outflow temperature, and outflow velocity. We compare our results in these tables to those in literature in the following subsections. For each exoplanet in this analysis, the posterior distributions of the model parameters we retrieve, as well as the transmission spectrum fits, can be found in the Appendix.

\begin{table}[]
\caption{Bounds on the peak excess absorption of the helium-triplet feature.} 
\begin{tabular}{lll}
\hline
\hline
Object    & $95$\% Bound & $99.7$\% Bound  \\
\hline
GJ-3470b  &$1.88$\%&$2.06$\%\\
HAT-P-11b &$1.32$\%&$1.42$\%\\
HD 189733b &$1.18$\%&$1.26$\%\\
HD 209458b &$0.83$\%&$0.90$\%\\
WASP-52b  &$4.06$\%&$4.42$\%\\
WASP-69b  &$3.38$\%&$3.49$\%\\
WASP-107b &$8.26$\%&$8.72$\%\\
WASP-177b &$1.18$\%&$1.46$\%\\
TOI-560b  &$0.67$\%&$0.74$\%\\
TOI-1430b &$0.68$\%&$0.75$\%\\
TOI-1683b &$0.70$\%&$0.88$\%\\
TOI-2076b &$1.06$\%&$1.12$\%\\
\end{tabular}
\label{tab:ObsResults}
\end{table}

\subsection{GJ-3470b}

\cite{Lampon2021} utilized a one-dimensional hydrodynamic model and a non-LTE model from \citet{Lampon2020} to estimate the mass-loss parameters for this planet, finding a temperature of the outflow of 5100$\pm$900K and a mass loss rate of $1.9 \pm 1.1 \times 10^{11}$ g s$^{-1}$. For H/(H + He) = $.90$, our results agree with \citet{Lampon2021}, falling inside of their outlined ranges. For H/(H + He) = $.99$,  our results agree well with \citet{Lampon2021}. \citet{Palle2020} used the modeling methods of \citet{Lampon2020} to bound the temperature and mass loss between $3 \times 10^{10}$ g s$^{-1}$ for $T = 6000$K and $10 \times 10^{10}$ g s$^{-1}$ for $T = 9000$K. These results differ from each other potentially due to \citet{Lampon2021} using Lyman-$\alpha$ observations to derive the neutral hydrogen density. The disagreement with \citet{Palle2020} for H/(H + He) = $.99$ likely results from the differing modeling framework, specifically the inclusion of short length-scale correlated noise we are modeling in this dataset. 

\subsection{HD 189733b}

We find that this planet has the highest outflow temperature in the entire sample, and even higher than those inferred by \citet{Lampon2021}, who report a temperature of $12400^{+400}_{-300}$K. \citet{Lampon2021} utilized the data from \citet{Salz2018} to infer mass-loss parameters. The authors report a mass-loss rate of $1.1 \pm 0.1 \times 10^{11}$ g s$^{-1}$. For a H/(H + He) value of $.99$, we retrieve a higher mass-loss rate and higher temperature when compared to \citet{Lampon2021}. There is large-amplitude correlated noise in this dataset (see Figure \ref{fig:hd189systematics}), which would make our modeling efforts diverge. Regardless, our results are consistent in identifying an extremely hot and large outflow from the hot Jupiter. It is also notable that in \citet{Fu2024}, the authors indicate that HD 189733b has a atmospheric metallicity that is three to five times stellar, meaning mass-loss modeling efforts that account for metals in the atmosphere may perform better and should be further studied \citep[e.g.,][]{linssen2024, Zhang24}.

\subsection{HD 209458b}

The \citet{Lampon2020} model was developed in a paper constraining the mass-loss parameters of this object, where they report a mass-loss and temperature range of $0.42 - 1.00 \times 10^{11}$ g s$^{-1}$ and 7125-8125K, respectively. Within the uncertainties, our results are consistent with \citet{Lampon2020}, but our model favors a lower temperature range. The dataset contains short length-scale correlated noise that we account for, which inflates our uncertainties on the retrieved parameters and makes tight parameter constraints difficult.

\subsection{HAT-P-11b}

In the original paper describing \textsf{p-winds}, \citet{pwinds} retrieved an escape rate of $2.5 \times 10^{10}$ g s$^{-1}$ and a temperature of 7200 K. As expected, this is consistent with our analysis when assuming solar helium abundance. \citet{Allart2018} employs EVE to set an upper bound of the metastable helium mass loss rate at $3 \times 10^5$ g s$^{-1}$ and a temperature normalized to mean atomic weight bound of $T/\mu \geq 24000 \text{ K} \cdot \text{amu}^{-1}$ . Our estimated upper bound of the metastable helium mass-loss rate of $8\times 10^4$ assuming solar helium abundance falls well within consistency with this result. When assuming H/(H + He) = $.99$, our upper bound on the metastable helium mass-loss rate is nearly identical to that of \citet{Allart2018}. 

Our temperature constraints of $7600^{+730}_{-670}$ K for $.90$ and $6500^{+620}_{-570}$ K for $.99$ correspond to $T/\mu = 5800^{+560}_{-510} \text{ K} \cdot \text{amu}^{-1}$ and $T/\mu = 6300^{+600}_{-490} \text{ K} \cdot \text{amu}^{-1}$ for $.90$ and $.99$ respectively.

The relatively large uncertainties in the transmission spectrum data points makes the constraints on the mass-loss parameters relatively wide and difficult to tightly constrain.

\subsection{WASP-52b}

Using \textsf{p-winds}, \citet{Kirk2022} used NIRSPEC observations to retrieve a mass-loss rate of $1.4 \times 10^{11}$ g s$^{-1}$ and outflow temperature of 8000K. Assuming a solar helium abundance, our escape rate is consistent with this result and our retrieved outflow temperature is lower, but within 3$\sigma$ of our estimate given the nearly 1000K errors on the temperature from both retrievals. Given the use of \textsf{p-winds} with solar composition, as \citet{Kirk2022} used, we would expect our sub-solar abundance result to be a mis-match. Furthermore, WASP-52b contains large-amplitude correlated noise in its helium transmission spectrum, meaning a comparison with a pure Parker wind model should differ. 

It should be noted that the relatively high escape rate of WASP-52b is paired with a relatively cool outflow temperature. HD 189733b has a similarly high escape rate, but a much hotter outflow. Future study should investigate the relationship between mass-loss rates and outflow temperatures and what could cause a high-escape rate paired with a cool outflow. It appears that the low density, extended planets roughly follow this trend of high escape rates and cool outflows.

\subsection{WASP-69b}

\citet{Vissapragada2020} utilizing a grid of models following the methods of \citet{Oklopcic2018} to find mass-loss rates and temperatures consistent with observations taken with the Wide-field Infrared Camera on the 200-inch Hale Telescope at Palomar Observatory. Under their assumption of solar-like hydrogen-to-helium abundances, our results do not fall in the grid of models consistent with these observations. The observational (photometric vs spectroscopic) and modeling differences in the studies could explain the variability in results, or there is some time variability in the helium signal from WASP-69b, perhaps due to the extended helium tail \citep{Nortmann2018}. 

\subsection{WASP-107b}

\citet{spake2018} utilized a one-dimensional Parker wind model from \citet{Oklopcic2018} to recover a mass-loss rate between $10^{10}$-$3\times10^{11}$ g s$^{-1}$, which is consistent with the results presented here. \citet{spake2018} utilized WFC3 onboard of \textit{Hubble Space Telescope} to observe the helium signature. \citet{Allart2019} used the EVaporating Exoplanet (EVE) code, finding the escape rate of metastable helium as $8 \times 10^5$ g s$^{-1}$, which is significantly larger than our upper-bound estimation of about $10^5$ g s$^{-1}$ for solar abundance. This becomes more consistent with our sub-solar helium abundance upper-bound estimation of about $2.5 \times 10^5$ g s$^{-1}$. \citet{Allart2019} utilized CARMENES at the Calar Alto observatory.

\subsection{WASP-177b}

Using \textsf{p-winds}, \citet{Kirk2022} used NIRSPEC observations to set an upper-limit on a mass-loss rate of $7.9 \times 10^{10}$ g s$^{-1}$ and outflow temperature of 6600~K. Our escape rate is consistent with this upper bound and our retrieved outflow temperature is within the error bounds on the temperature estimate for both abundance assumptions. The significant amount of systematics within the NIRSPEC observation makes the constraints very difficult to obtain. 

\subsection{The Mini-Neptunes}

For the objects TOI-560b, TOI-1430b, TOI-1683b, and TOI-2076b, \citet{Zhang_2023} utilizes an order of magnitude estimate outlined in \citet{Zhang_2022} and the Parker wind model outlined in \citet{Oklopcic2018} to constrain the mass-loss rates of the three mini-Neptunes. Mini-Neptunes could plausibly have high metallicity atmospheres, which inhibits the ability of Parker wind models to accurately constrain mass-loss parameters without a constraint on the atmospheric metallicity. 

There are discrepancies when comparing our results to the results in \citet{Zhang_2023}, likely due to the handling of systematics. Each of the young planets have significant correlated noise features that the Gaussian Process fit, meaning that the resulting mass-loss rate and outflow temperatures will be affected when comparing to other Parker wind models. Modeling differences are the most-likely cause of discrepancies between our results and those described in \citet{Zhang_2023}. The inclusion of Gaussian processes with datasets containing correlated noise would cause deviations in results, even when comparing two Parker-wind models.

\subsubsection{TOI-560b}

\citet{Zhang_2023} reports mass-loss rates of $1.6 \times 10^{11}$ g s$^{-1}$ and $2.1 \times 10^{10}$ g s$^{-1}$ for the Parker and order of magnitude estimates, respectively. The authors also find a median outflow temperature of 9800 K using the Parker wind model. The Parker wind estimate that the authors present are consistent with both of our helium abundance values, but the sub-solar abundance retrieval's outflow temperature is much more suppressed than that presented in \citet{Zhang_2023}. TOI-560b contains large-amplitude correlated noise features which impacted fits.

\subsubsection{TOI-1430b}

\citet{Zhang_2023} reports mass-loss rates of $1.3 \times 10^{11}$ g s$^{-1}$ and $2.7 \times 10^{10}$ g s$^{-1}$ for the Parker and order of magnitude estimates, respectively. The authors also find a median outflow temperature of 6700 K using the Parker wind model. TOI-1430b has large-amplitude and short-length-scale correlated noise that impacts the fitting of an atmospheric escape model.

\subsubsection{TOI-1683b}

\citet{Zhang_2023} reports mass-loss rates of $2.5 \times 10^{10}$ g s$^{-1}$ and $2.2 \times 10^{10}$ g s$^{-1}$ for the Parker and order of magnitude estimates, respectively. The authors also find a median outflow temperature of 6700 K using the Parker wind model. TOI-1683b contains relatively well-behaved correlated noise, but the magnitude of the helium triplet is suppressed by an offset that the GP corrected for, meaning our mass loss rates are understandably suppressed when comparing the solar-abundance retrievals.

\subsubsection{TOI-2076b}

\citet{Zhang_2023} reports mass-loss rates of $2.4 \times 10^{10}$ g s$^{-1}$ and $2.8 \times 10^{10}$ g s$^{-1}$ for the Parker and order of magnitude estimates, respectively. The authors also find a median outflow temperature of 5000 K using the Parker wind model. TOI-2076b, similarly to TOI-1430b, has large-amplitude and short length-scale correlated noise that impacts the fitting of an atmospheric escape model.

\section{Interpretation}\label{sec:interpretation}

Given the utilization of a new modeling technique in metastable helium triplet analyses in the form of Gaussian Processes, we wish to interpret both the individual modeling efforts and the results of the population-level modeling results. Thus, this section is split into two subsections, Section \ref{sec: interp-GP} will discuss our introduction of GPs into the modeling efforts and Section \ref{sec: interp-results} will discuss the interpretations of the model results. 

\subsection{Gaussian Processes}\label{sec: interp-GP}

In Section \ref{sec: GP}, we discuss the intricacies of our efforts to mediate the correlated noise in the transmission spectra via Gaussian Processes. One important question to answer is how the GP affects our model results. We present an example of systematics within a helium observation and the ability of GPs to fit for the systematics. Figure \ref{fig:hd189systematics} exemplifies the impacts of systematics within the helium-triplet window. 

We use HD 209458b and WASP-69b as examples of large- and small-amplitude systematics within the datasets to investigate the difference in model fits accounting for correlated noise fits and not. To compare model fits, we utilize Bayesian evidences to investigate whether the GPs are a preferred fit and look at the posterior distributions for any differences. Table \ref{tab:nogpcompare} outlines these evidences and posteriors.

\begin{table}[ht!]
\centering
\caption{Comparisons of the Bayesian evidences and posteriors of the WASP-69b and HD 209458b retrievals with and without GPs. We assume $.99$ for the hydrogen-to-helium ratio. The uncertainties are reported at the 16th and 84th percentiles.} 
\begin{tabular}{lllll}
\hline
\hline
GP-Model       & $\log{\dot{m}}$ & $\log{T}$ & $v$ & $\log{\mathcal{Z}}$  \\
\hline
WASP-69b  &$11.34^{+0.04}_{-0.04}$&$3.67^{+0.01}_{-0.01}$&$-2.3^{+0.4}_{-0.4}$&$1967$\\
HD 209458b &$11.12^{+0.42}_{-0.85}$&$3.78^{+0.11}_{-0.15}$&$+4.2^{+0.9}_{-1.1}$&$5247$\\
\hline
\hline
No-GP Model    & $\log{\dot{m}}$ & $\log{T}$ & $v$ & $\log{\mathcal{Z}}$  \\
\hline
WASP-69b  &$11.35^{+0.03}_{-0.03}$&$3.67^{+0.01}_{-0.01}$&$-2.3^{+0.2}_{-0.2}$&$1954$\\
HD 209458b &$11.57^{+0.10}_{-0.12}$&$3.89^{+0.03}_{-0.03}$&$+3.4^{+0.6}_{-0.6}$&$5069$\\ \hline
\end{tabular}
\label{tab:nogpcompare}
\end{table}

For objects with minimal correlated noise in their transmission spectrum, such as WASP-69b, we observe that the posterior distributions are nearly identical, but with a slight inflation of the uncertainties. This is to be expected, as Gaussian processes provide more realistic uncertainties on retrieved values. The log-evidences favor the GP model significantly.

For objects with moderate correlated noise in their transmission spectrum, such as HD 209458b, we observe that the posterior distributions diverge, but with a similar inflation of the uncertainties. The posteriors are consistent with one another, but we would expect the posteriors when including GPs to be more accurate in retrieving mass-loss parameters. The log-evidences significantly favor the GP model, more so than for datasets with less correlated noise.

Including GPs in our mass-loss parameter retrievals allows us to present uncertainties from a model that is fitting for complex correlated noise sources allowing for more accurate retrieved values with wider uncertainties that capture the true variation due to this correlated noise. The evidences presented strongly prefer fits with the GPs and we can be more confident in the retrieved median values for the mass-loss parameters.

We present the \textsf{p-winds} fitted transmission spectra with and without GPs being used. The transmission spectrum of HD 209458b has larger correlated noise relative to the depth of the helium-triplet line, and thus displays a difference in the fitted model when compared to WASP-69b, which has a relatively small amount of correlated noise. In the case of WASP-69b, the two models are essentially indistinguishable visibly. 

\begin{figure}[ht!]
    \centering
    \includegraphics[width=.45\textwidth]{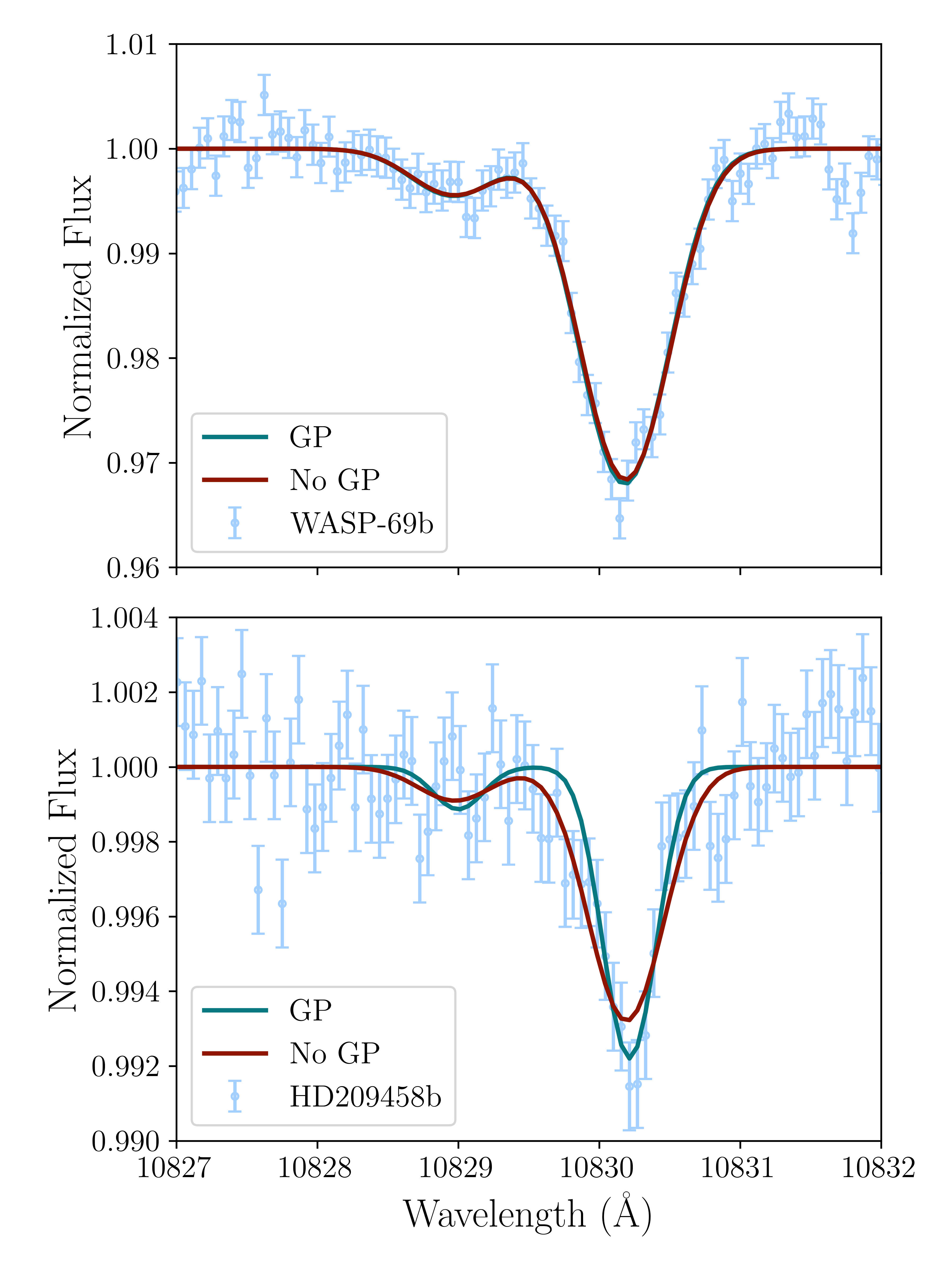}
    \caption{Comparative figures of models with and without Gaussian Processes included. The line in green  represents the \textsf{p-winds} fit with the GP included. The line in red represents the \textsf{p-winds} fit without a GP included. H/(H + He) = $.99$.}
    \label{fig:comparison}
\end{figure}

It should be expected that the difference between the two models are larger for spectra that have more correlated noise, as is the case with HD 209458b. We conclude that the GP is valuable for fitting correlated noise and producing more realistic uncertainties (now accounting for uncertainties from correlated noise) in the retrieved parameters.

\subsection{Model Interpretation}\label{sec: interp-results}

The energy-limited mass loss formulation was initially theorized by \citet{Watson1981} to model the evolution of early Earth and Venus, which had H-rich envelopes in the beginning of their lives. More recently, this formulation has routinely been used to model the evolution of exoplanets \citep[e.g.,][]{Lecavelier2007, Kurokawa2013, Lopez2014, Owen2017, Mordasini2020}. This theory poses that the XUV irradiation from the host star is absorbed in the upper atmosphere of a given planet, which is converted to heat, causing the gas to expand and producing an outflow. The conversion between stellar radiation into outflow is a complex process and is usually studied in the form of an efficiency factor.

\citet{Vissapragada2022b} presented a mean energy-limited outflow efficiency for a restricted population of exoplanets on close-in orbits using photometry. The authors conclude that the retrieved efficiency of photoevaporation is too low to carve out the upper boundary of the Neptune desert. With our population of 12 exoplanets with detections from spectroscopy, we repeat the analysis to obtain an energy-limited outflow efficiency, with the inclusion of hot Jupiters and mini-Neptunes. The energy-limited approximation from \citet{Caldiroli2022} states:

\begin{equation}
    \dot{M} = \frac{3 \varepsilon}{4G}\frac{F_{XUV}}{\rho_p}\mathrm{,}
\end{equation}

where $\varepsilon$ is the efficiency with which XUV photons are converted into escaping material, $G$ is the gravitational constant, $F_{XUV}$ is the incident XUV flux in the upper atmosphere and $\rho_p$ is the planetary density. The efficiency $\varepsilon$ factor encapsulates different processes that can potentially decrease or increase the planetary mass-loss rate \citep{Shematovich2014}, such as radiative cooling \citep{MClay2009}, interactions with the stellar wind \citep{Carolan2021}, magnetic field interactions \citep{BJaffel2022}, radiation pressure \citep{Bourrier2016}, and Roche lobe overflow \citep{Erkaev2007}.

\begin{figure*}[ht!]
    \centering
    \includegraphics[width=\textwidth]{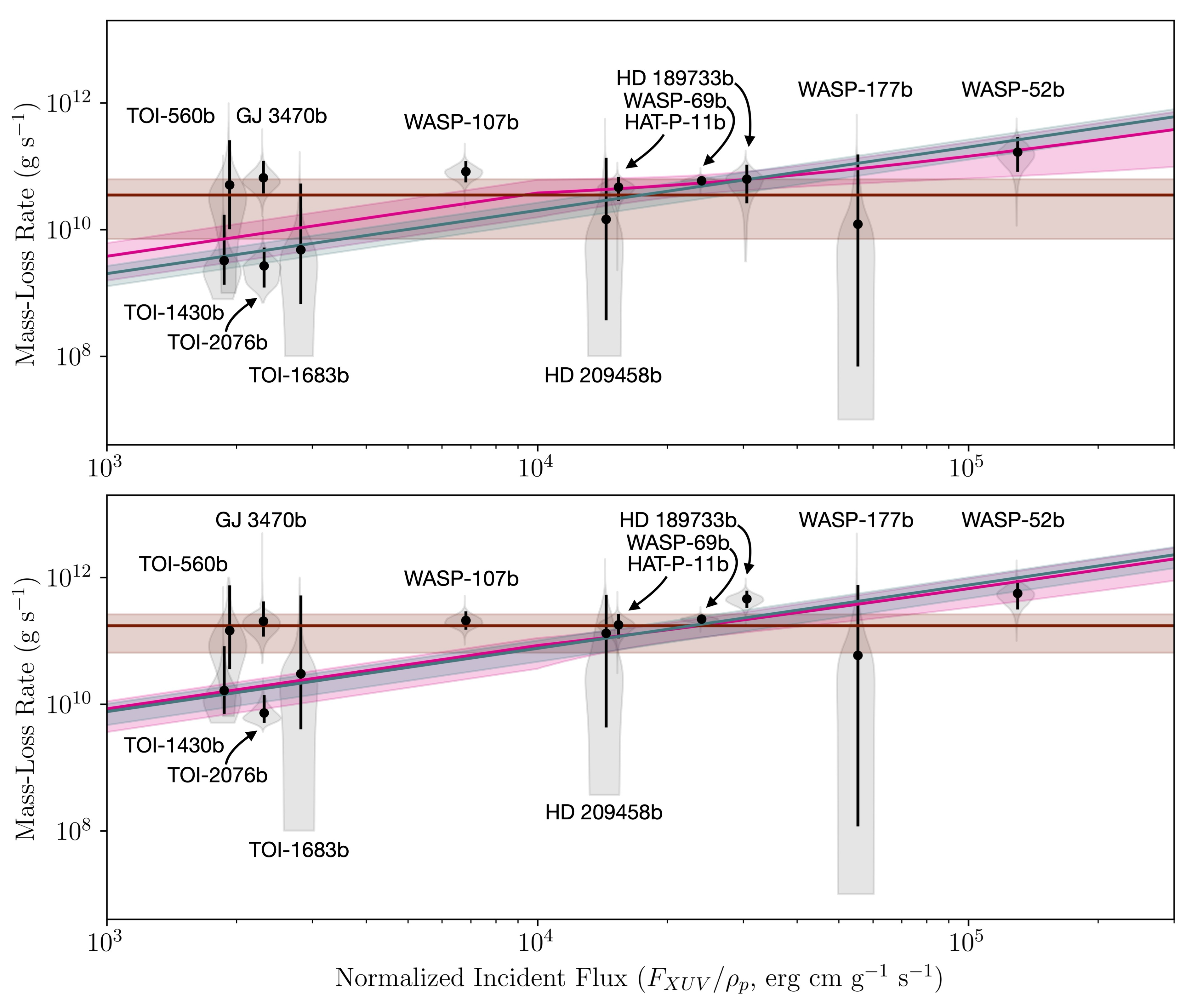}
    \caption{Mass-loss rate, $\dot{M}$, as a function of the normalized incident flux, $F_{XUV}/\rho$, for the sample assuming a H/(H + He) abundance of $.90$ (top panel) and $.99$ (bottom panel). The violin plots in grey illustrate the marginalized distribution of mass-loss rates with the median, 5\% lower bound, and 95\% upper bound overplotted on the violin plots in black. The dark green line illustrates the best-fit line for mass-loss rate as a function of the normalized incident flux. The pink line illustrates the best-fit piecewise model for mass-loss rate as a function of the normalized incident flux. The dark red line illustrates the fitted flat line.}
    \label{fig:fxuv_mdot}
\end{figure*}

Theory expects a linear relationship between the ratio of $F_{XUV}/\rho_p$ and the mass-loss rate, $\dot{M}$. By fitting a line with the retrieved mass-loss rates and the stellar/planet properties outlined above, we obtain Figures \ref{fig:fxuv_mdot}.

Beyond $F_{XUV}/\rho_p \sim 10^4$, this specific formulation breaks down \citep{MClay2009}; we also fit a piecewise line in the energy-limited regime, then a separate line beyond the energy-limited regime threshold. Finally, we fit a flat line to the data for a baseline.

\subsubsection{Pure Energy-Limited Mass Loss}

\citet{Caldiroli2022} describes a relationship between the normalized XUV flux and the mass-loss rate, mediated by the efficiency of atmospheric mass loss, denoted as $\varepsilon$. This efficiency reflects how effectively an exoplanet’s atmosphere loses mass in response to increasing XUV irradiation.

For $\text{H}/(\text{H + He}) = .90$, we applied a linear fit to the exoplanet population data, resulting in a mass-loss efficiency of $0.18 \pm 0.04$. This value represents the average efficiency across the sample. For a hydrogen-to-helium ratio of $.99$, the linear fit yielded a higher efficiency of $0.67 \pm 0.14$. As expected, more hydrogen rich escape models indicate higher efficiencies of converting XUV irradiation into mass loss.

A visual inspection of Figures \ref{fig:fxuv_mdot} suggest that there may be an ambiguous correlation between increasing XUV flux and the mass-loss rate for the exoplanets in our sample. To explore this, we fit a flat-line model, a purely linear, and a linear model that changes slope at $F_{XUV}/\rho_p \sim 10^4$. We hypothesize that the mass-loss efficiency $\varepsilon$ might effectively be zero and seek for evidence that energy-limited mass loss is supported by these results.

The assumption of energy-limited mass loss begins to break down at $F_{XUV}/\rho_p \sim 10^4$, where the conversion of XUV flux into mass loss becomes less efficient, as suggested by \citet{Vissapragada2020}. To test this, we fit the data using a piecewise linear model, which introduces a change in slope at this threshold.

For a hydrogen-to-helium ratio of $.90$, the efficiency before the boundary was $0.34 \pm 0.13$, which dropped to $0.10 \pm 0.06$ beyond the threshold. For a hydrogen-to-helium ratio of $.99$, the efficiency prior to the boundary was higher at $0.75 \pm 0.21$, decreasing to $0.57 \pm 0.19$ post-boundary. The results indicate a significant reduction in XUV-conversion efficiency beyond this boundary. \citet{Vissapragada2022b} constrained the mean energy-limited outflow efficiency to $\varepsilon = .41^{+.16}_{-.13}$, consistent with our solar-abundance efficiency pre-boundary.

When comparing the models, for an abundance of $.90$, the log evidence values for the flat line, linear, and piecewise models were $-311.69$, $-306.55$, and $-306.36$, respectively, indicating a preference for models that account for energy-limited mass loss. For an abundance of $.99$, the log evidence values for the flat line, linear, and piecewise models were $-325.58$, $-321.11$, and $-321.23$, respectively. Here too, the data support the energy-limited mass-loss models. No matter the abundances used, our exoplanet population favors strong evidence of energy-limited mass loss models. See \citet{Benneke_2013} for log-evidence model selection criteria, where a Bayes Factor is matched with a significance value. Table \ref{tab:Results9099} contains the full results of the runs for each abundance ratio used.

\begin{table*}[]
\centering
\caption{Results of the \textsf{p-winds} retrievals outlining each exoplanet's mass-loss rate $\dot{m}$, outflow temperature $T$, and outflow velocity, $v$, assuming H/(H + He) abundances of $.99$ and $.90$.}
\begin{tabular}{l|ccc|ccc}
\hline
\hline
Object    & \multicolumn{3}{c|}{Abundance = $.99$} & \multicolumn{3}{c}{Abundance = $.90$} \\
          & $\dot{m}$ (g s$^{-1}$) & $T$ (K) & $v$ (km s$^{-1}$) & $\dot{m}$ (g s$^{-1}$) & $T$ (K) & $v$ (km s$^{-1}$) \\
\hline
GJ-3470b  & $2.0^{+0.8}_{-0.6} \times 10^{11}$ & $4500^{+320}_{-300}$ & $-3.0^{+1.6}_{-1.6}$ & $6.6^{+2.9}_{-1.9} \times 10^{10}$ & $6600^{+810}_{-580}$ & $-2.9^{+1.6}_{-1.6}$ \\
HAT-P-11b & $1.8^{+0.5}_{-0.4} \times 10^{11}$ & $6500^{+620}_{-570}$ & $-1.8^{+0.9}_{-0.9}$ & $4.7^{+1.2}_{-1.1} \times 10^{10}$ & $7600^{+730}_{-670}$ & $-1.8^{+0.9}_{-0.9}$ \\
HD 189733b & $4.6^{+0.9}_{-0.8} \times 10^{11}$ & $15000^{+700}_{-670}$ & $-2.9^{+0.9}_{-0.9}$ & $6.3^{+2.4}_{-2.2} \times 10^{10}$ & $17000^{+2100}_{-1500}$ & $-2.9^{+0.9}_{-0.9}$ \\
HD 209458b & $1.3^{+2.1}_{-1.1} \times 10^{11}$ & $6000^{+1700}_{-1800}$ & $+4.2^{+0.9}_{-1.1}$ & $1.4^{+5.5}_{-1.3} \times 10^{10}$ & $6800^{+2600}_{-1900}$ & $+3.5^{+1.2}_{-1.2}$ \\
WASP-52b  & $5.6^{+2.0}_{-1.6} \times 10^{11}$ & $4600^{+560}_{-500}$ & $-2.2^{+1.1}_{-1.0}$ & $1.7^{+0.7}_{-0.5} \times 10^{11}$ & $5900^{+1000}_{-990}$ & $-2.3^{+1.1}_{-1.1}$ \\
WASP-69b  & $2.2^{+0.2}_{-0.2} \times 10^{11}$ & $4700^{+110}_{-110}$ & $-2.3^{+0.4}_{-0.4}$ & $5.9^{+0.6}_{-0.5} \times 10^{10}$ & $6200^{+140}_{-140}$ & $-2.3^{+0.4}_{-0.4}$ \\
WASP-107b & $2.1^{+0.5}_{-0.4} \times 10^{11}$ & $2500^{+120}_{-110}$ & $-4.9^{+0.6}_{-0.7}$ & $8.3^{+2.2}_{-1.7} \times 10^{10}$ & $3600^{+260}_{-240}$ & $-4.8^{+0.6}_{-0.7}$ \\
WASP-177b & $5.9^{+26}_{-5.8} \times 10^{10}$ & $4900^{+6300}_{-2800}$ & $+6.0^{+4.2}_{-5.4}$ & $1.2^{+6.2}_{-1.2} \times 10^{10}$ & $5600^{+5300}_{-2900}$ & $+6.5^{+3.8}_{-4.3}$ \\
TOI-560b  & $1.5^{+2.6}_{-0.9} \times 10^{11}$ & $5000^{+1400}_{-1000}$ & $+4.0^{+2.9}_{-3.5}$ & $5.1^{+8.7}_{-3.0} \times 10^{10}$ & $6900^{+3300}_{-1900}$ & $+4.0^{+3.1}_{-4.1}$ \\
TOI-1430b & $1.7^{+2.6}_{-0.8} \times 10^{10}$ & $2500^{+580}_{-370}$ & $-4.3^{+1.5}_{-1.8}$ & $3.2^{+5.9}_{-1.6} \times 10^{9}$ & $2600^{+840}_{-490}$ & $-3.8^{+1.4}_{-1.7}$ \\
TOI-1683b & $3.0^{+14}_{-2.3} \times 10^{10}$ & $3500^{+2200}_{-1000}$ & $-6.0^{+3.6}_{-3.3}$ & $4.8^{+17}_{-3.6} \times 10^{9}$ & $3700^{+2300}_{-1200}$ & $-6.1^{+3.1}_{-3.0}$ \\
TOI-2076b & $7.2^{+3.5}_{-1.6} \times 10^{9}$ & $1300^{+270}_{-150}$ & $-1.0^{+0.6}_{-0.6}$ & $2.6^{+1.4}_{-1.0} \times 10^{9}$ & $1600^{+280}_{-270}$ & $-1.2^{+0.6}_{-0.6}$ \\
\hline
\end{tabular}
\label{tab:Results9099}
\end{table*}

\begin{figure}[ht!]
    \centering
    \includegraphics[width=.45\textwidth]{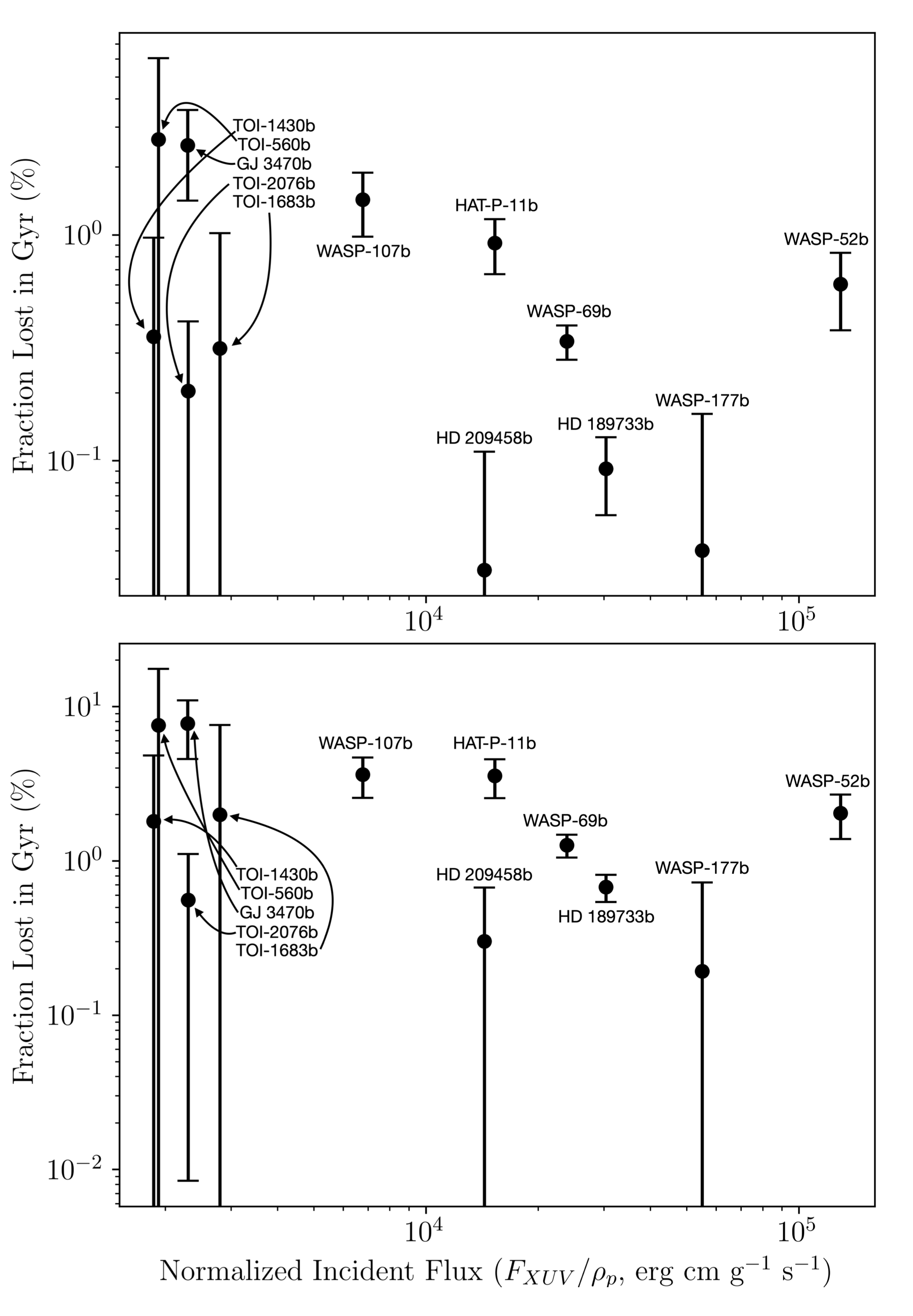}
    \caption{Fraction of mass lost for each target in 1 Gyr as a function of the normalized incident flux, $F_{XUV}/\rho$, for the sample assuming a H/(H + He) abundance of $.90$ (top panel) and $.99$ (bottom panel).}
    \label{fig:masslossgyr90}
\end{figure}

Paramount to the motivation of this work was questioning whether small, highly-irradiated planets are able to retain their atmospheres. To investigate this, we utilize the retrieved mass-loss rates to estimate the fraction of total mass that a planet loses in a Gyr. Figures \ref{fig:masslossgyr90} illustrate this assuming a constant mass-loss rate throughout the lifetime of the exoplanet.

Regarding the solar-like hydrogen abundance mass-loss rates, it appears that small objects, such as the mini-Neptunes, cannot hold onto their atmospheres for a significant amount of time, and thus evolve into super-Earths. However, the mass-loss rates must be better constrained to provide a complete picture. The mini-Neptunes could all plausibly lose a percent of their atmosphere in a Gyr, which would be a significant portion of their atmospheres. TOI-2076b is the exception to this, losing much less than 1\% of its atmosphere in a Gyr. 

GJ-3470b and HAT-P-11b are warm Neptunes, while WASP-107b is an inflated warm Jupiter with an extended tail. The results indicate that warm Neptune atmospheric evolution is significantly impacted by photoevaporative effects, while hot Jupiters are robust against it, save those with extremely low densities and extended tails. Mini-Neptunes appear to have unique evolutionary paths that evolve them into super Earths, as indicated in the literature \citep{Zhang_2023}. 

For a sub-solar helium abundance, all targets lose significantly more mass than their solar-abundance counterparts. Each object could plausibly lose 1\% of its mass. For warm Neptunes like HAT-P-11b and GJ-3470b, a near 10\% loss in total mass would constitute a large fraction of its atmosphere. This is less prominent for the hot Jupiters, with higher mass content being contained in their atmospheres. WASP-69b and WASP-52b have extended atmospheres and would be expected to have high mass-loss rates, and thus high fractional mass loss. WASP-107b's extreme low density also explains its high fractional mass loss relative to its hot Jupiter counterparts.

Constraining the true hydrogen and helium abundances in exoplanet atmospheres is critical to our understanding of atmospheric evolution, leading to differing understandings of how planets will evolve in their lifetimes. While the community is beginning to understand that super-solar hydrogen abundances are better applicable to exoplanet atmospheres \citep{Lampon2021}, it is important to fully understand nuances and subtleties regarding atmospheric abundances.

\section{Conclusions}\label{sec:conclusions}

In this work, we performed a uniform mass-loss analysis of all publicly available spectroscopic helium triplet observations, allowing us to examine population trends in atmospheric escape among exoplanets. Atmospheric mass loss is pivotal to understanding exoplanet evolution, yet the efficiency with which XUV irradiation is converted into mass loss has remained ambiguous due to inconsistencies in previous analyses. \citet{Caldiroli2022} outlined a theoretical relationship between an exoplanet’s mass-loss rate and XUV irradiation, governed by an efficiency factor under the assumption that the planet is in the energy-limited mass-loss regime (up to $F_{\text{XUV}}/\rho_p \sim 10^4$). Using the open-source \textsf{p-winds} model, Gaussian processes, and nested sampling, we constrained this efficiency for two different hydrogen fractions, $.90$ and $.99$, finding values of $\varepsilon$ = $0.34$ $\pm$ $0.13$ for the $.90$ ratio and $\varepsilon$ = $0.75$ $\pm$ $0.21$ for the $.99$ ratio.

We conclude that the current sample of helium triplet observations provides strong evidence supporting energy-limited mass loss. While additional observations are necessary to refine population trends, our results lend credence to the theoretical framework underlying energy-limited escape. The observational capabilities of \textit{JWST} offer a unique opportunity to obtain more precise mass-loss estimates across a broader range of exoplanet populations \citep{dossantos2023}. We also call for the publication of non-detections of the helium triplet signature. While non-detections can only provide upper bounds on mass-loss parameters, these bounds are critical for constraining our understanding of planetary evolution. Each non-detection contributes valuable information to population-level studies of evaporating exoplanets. 

It is also essential to direct future studies toward a deeper understanding of the host stars of exoplanets undergoing mass loss. A fundamental input for mass-loss models is the stellar spectral energy distribution (SED), which remains unconstrained for many of the systems in this sample. Although using SED proxies is a common approach, this method introduces uncertainties that are challenging to quantify. Expanding observations to include exoplanets orbiting a wider range of stellar types will help elucidate how mass-loss rates vary across non-solar-type stars.

One significant limitation of current helium observations is the bias toward detecting large helium signatures, and thus higher mass-loss rates. Theoretical models of mass loss do not inherently favor high or low rates, meaning the population trends outlined in this study primarily apply to exoplanets with substantial mass loss. Observations of exoplanets with moderate to slow mass loss are technically challenging and difficult to secure telescope time for, but are essential for fully understanding mass-loss population trends.

In our analysis, we identified planets that exhibit high mass-loss rates, but relatively cool outflow temperatures. Although these objects seem to be low density, inflated objects, it is worth an investigation into the relationship between mass-loss rates and outflow temperatures to understand the physics of this trend. 

With our observational evidence of energy-limited mass loss, we now have a clearer understanding of the dominant physical processes affecting highly irradiated exoplanets. This enhances our ability to contextualize key aspects of atmospheric evolution, such as the exoplanetary radius gap. However, as \citet{Vissapragada2022b} concluded, energy-limited outflows alone may be too inefficient to account for the upper edge of the Neptune desert. They suggest that additional mechanisms, such as tidal disruption, may be required to fully explain certain features of planetary evolution \citep{owen18}.

With the growing number of helium triplet observations and advances in observational capabilities, we are poised to deepen our understanding of atmospheric escape mechanisms. Continued exploration will not only clarify the role of energy-limited mass loss, but also provide a more complete picture of exoplanetary evolution across a diverse range of systems.

\begin{acknowledgments}
The authors would like to thank Jorge Sanz-Forcada for helpful insight regarding XUV-flux calculations as well as sharing the high-energy SEDs of HD~189733 and HD~209458. This research received support through the program HST-AR-17035 provided by NASA as a grant from the Space Telescope Science Institute (STScI). STScI is operated by the Association of Universities for Research in Astronomy, Inc. under NASA contract NAS 5-26555. STScI stands on the traditional and unceded territory of the Piscataway-Conoy and Susquehannock peoples. We acknowledge the often-overlooked labor of the custodial, facilities and security staff at STScI -- this research would not be possible without them. This research made use of the NASA Exoplanet Archive, which is operated by the California Institute of Technology, under contract with the National Aeronautics and Space Administration under the Exoplanet Exploration Program. This research used data from the CARMENES data archive at CAB (CSIC-INTA). The CARMENES archive is part of the Spanish Virtual Observatory project (http://svo.cab.inta-csic.es), funded by MCIN/AEI/10.13039/501100011033/ through grant PID2020-112949GB-I00. Some of the data presented herein were obtained at the W. M. Keck Observatory, which is operated as a scientific partnership among the California Institute of Technology, the University of California and the National Aeronautics and Space Administration. The Observatory was made possible by the generous financial support of the W. M. Keck Foundation. The authors wish to recognize and acknowledge the very significant cultural role and reverence that the summit of Maunakea has always had within the indigenous Hawaiian community. We are most fortunate to have the opportunity to conduct observations from this mountain.
\end{acknowledgments}

\vspace{5mm}
\facilities{CAO:3.5m (CARMENES), Keck:II (NIRSPEC)}

\software{\textsf{astropy} \citep{2013A&A...558A..33A,2018AJ....156..123A,2022ApJ...935..167A}, \textsf{p-winds} \citep{pwinds}, \textsf{scipy} \citep{scipy}, \textsf{numpy} \citep{numpy}, \textsf{matplotlib} \citep{matplotlib}, \textsf{dynesty} (\citep{dynesty1}, \citep{dynesty2}, \citep{dynesty3}, \citep{dynesty4}, \citep{dynesty5}, \citep{dynesty6}), \textsf{celerite} \citep{celerite}, \textsf{ray} \citep{ray}, \textsf{pickle} \citep{pickle}, \textsf{corner} \citep{corner}, \textsf{pandas} \citep{pandas}, \textsf{timeout\_decorator}, \textsf{jupyter} \citep{jupyter}.}

\clearpage
\bibliography{AAS60118_new_v2.ms}{}
\bibliographystyle{aasjournal}

\appendix

We present here the model-parameter posteriors as well as the model-fit transmission spectra. 

The appendix figure sets are accessible on Zenodo \citep{mccreery_2025}.

\end{document}